\documentclass[12pt,fleqn,epsfig]{article}

\usepackage{epsfig}
\usepackage{fleqn,espcrc1}
\usepackage{wrapfig}

\newcommand{\beq}{\begin{equation}}
\newcommand{\eeq}{\end{equation}}
\newcommand{\bqa}{\begin{eqnarray}}
\newcommand{\eqa}{\end{eqnarray}}

\begin{document}
\title{Hard Thermal Loops and QCD Thermodynamics
\footnote{ Talk given at Conference on Strong and Electroweak
Matter (SEWM 2002), Heidelberg, Germany, October 2-5 2002.}}

\author{Jens O. Andersen\\
Institute for Theoretical Physics, University of Utrecht,\\
       Leuvenlaan 4, 3584 CE Utrecht, The Netherlands\\}

%\small

\maketitle
\abstract{The conventional
weak-coupling expansion for thermodynamic quantities in hot field theories
shows poor convergence unless the coupling constant is tiny.
I discuss screened perturbation theory (SPT)
which is a way of reorganizing the perturbative expansion for 
scalar theories %by adding and subtracting a local mass term in the Lagrangian,
and hard-thermal-loop perturbation theory (HTLPT), which is
its generalization to gauge theories.
I present results for the pressure to three loops in SPT and to two loops
in HTLPT. We compare the latter with three- and four-dimensional
lattice simulations of pure-glue QCD.

\section{Introduction}
The heavy-ion collision experiments at RHIC and LHC give us for the first time
the possibility to study the properties of the high-temperature phase
of QCD. There are many methods that can be used to calculate the
properties of a quark-gluon plasma. One of these methods is lattice gauge
theory, which gives reliable results for equilibrium properties such as the
pressure but cannot easily be applied to real-time processes.
Another method is the weak-coupling expansion, which can be applied to
both static and dynamical quantities. In the case of the pressure,
the weak-coupling expansion has been carried out to order 
$g^5$ for massless $\phi^4$ theory~\cite{az,ps,bns}, 
for QED~\cite{pp,a}, and for 
nonabelian gauge theories~\cite{az,kz,bnq}.
However, it turns out that the weak-coupling expansion 
does not converge unless the strong coupling constant $\alpha_s$ is tiny.
For instance the $g^3$ term is smaller than the $g^2$ term only if 
$\alpha_s\simeq1/20$. This corresponds to a temperature of $10^5$GeV, 
which is several orders of magnitude 
larger than those relevant for experiments at RHIC and LHC.

In this talk, I will discuss recent advances in the calculation of
thermodynamic quantities in hot field theories based on SPT and HTLPT.
Due to lack of time, I cannot critically compare the approach presented
here and other resummation methods that have recently been advocated 
by Blaizot, Iancu and Rebhan~\cite{BIR-99} and by Peshier~\cite{Peshier-00}
based on the phi-derivable approach~\cite{ward}
and hard thermal loops. Instead, I refer to the
talk at SEWM 2002 by A. K. Rebhan~\cite{proc}.

\section{Thermal scalar field theory}

Consider a massless scalar field theory with a quartic interaction.
The Euclidean Lagrangian is 
\begin{center}
\bqa
{\cal L}&=&{1\over2}(\partial_{\mu}\phi)^2+{g^2\over24}\phi^4
+\Delta{\cal L}\;,
\label{lorg}
\eqa
\end{center}
where $\Delta{\cal L}$ includes the counterterms needed to remove ultraviolet
divergences. At finite temperature, the naive perturbative expansion
in $g^2$ also generates infrared divergences. These infrared divergences can
be removed by resummation of the higher order diagrams that generate a
thermal mass. The resummed series is then an expansion in $g$ rather than
in $g^2$. Through order $g^5$ it reads~\cite{az,ps,bns}:
\bqa\nonumber
{\cal P} &=& {\cal P}_{\rm ideal} \left[
1-{5\over4}\alpha+{5\sqrt{6}\over3}\alpha^{3/2}+{15\over4}
\left(\log{\mu\over2\pi T}+0.40\right)\alpha^2\right.\\ 
&&\left.-{15\sqrt{6}\over2}\left(\log{\mu\over2\pi T}-{2\over3}\log\alpha
-0.72\right)\alpha^{5/2}+{\cal O}(\alpha^3\log\alpha)\right]\;,
\eqa
where ${\cal P}_{\rm ideal} = (\pi^2/90)T^4$
is the pressure of an ideal gas of free massless bosons,
$\alpha=g^2(\mu)/16\pi^2$, and $g(\mu)$ is the 
$\overline{\rm MS}$ coupling constant at the renormalization scale $\mu$.

In Fig.~\ref{fig1}, we show the successive perturbative 
approximations
to ${\cal P}/{\cal P}_{\rm ideal}$ as a function of $g(2\pi T)$.
The bands are obtained by varying the renormalization scale $\mu$ from 
$\pi T$ to $4\pi T$. In order to express $g(\mu)$ in terms of $g(2\pi T)$,
we solve the renormalization group equation for the running coupling
constant with a five-loop beta function. The poor convergence
of the weak-coupling expansion is evident from Fig.~\ref{fig1}.
The successive approximations fluctuate wildly and the bands become
broader with the increasing coupling $g$. This indicates large theoretical
errors.

\begin{wrapfigure}[]{l}[0pt]{8cm}
%\begin{figure}[htb]
\epsfysize=6cm
\epsffile{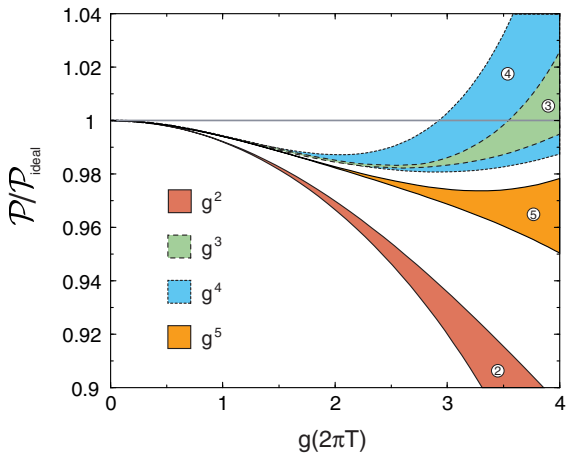}
%\vspace{2mm}
\caption[a]{
Weak-coupling expansion to orders $g^2$, $g^3$, $g^4$, and $g^5$
for ${\cal P}/{\cal P}_{\rm ideal}$ as a function of
$g(2\pi T)$.
}
\label{fig1}
\vspace{-1.cm}
%\end{figure}
\end{wrapfigure}
\subsection{Screened Perturbation theory}
Screened perturbation theory is a way of reorganizing perturbation
theory by adding and subtracting a local mass term in the Lagrangian.
It was introduced by Karsch, Patk\'os and Petreczky~\cite{KPP-97}
and can be made more systematic by using a framework called 
``optimized perturbation theory'' that was applied to a 
spontaneously broken scalar field theory~\cite{CK-98}.

The Lagrangian is written as
\bqa\nonumber
{\cal L}&=&{\cal E}_0+{1\over2}(\partial_{\mu}\phi)^2
+(1-\delta)m^2
+{g^2\over24}\phi^4
\\
&&
+\Delta{\cal L}+\Delta{\cal L}_{\rm SPT}\;,
\eqa
where ${\cal E}_0$ is the vacuum energy term and $m$ is a mass term. 
If we set $\delta=0$, 
we recover the original Lagrangian~(\ref{lorg}).
SPT is defined by treating $\delta$ to be of order $g^2$ and expanding
systematically in powers of $g^2$. The reorganization of the pertubative
series generates new ultraviolet divergences that are cancelled by
the addiational counterterms in $\Delta{\cal L}_{\rm SPT}$.

\begin{wrapfigure}[]{l}[0pt]{8cm}
%\begin{figure}[htb]
\epsfysize=6cm
\epsffile{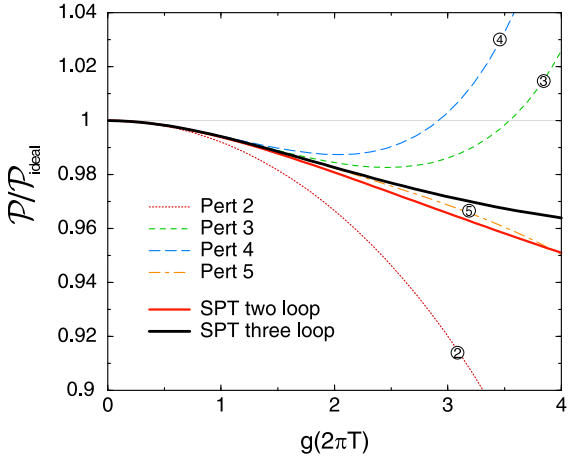}
%\vspace{2mm}
\caption[a]{The two- and three-loop approximations to the pressure
within SPT. For comparision, we have also included the sucessive
weak-coupling approxmations.}
\label{fig2}
\vspace{-1.8cm}

%\end{figure}
\end{wrapfigure}

\subsection{Mass prescriptions}
I would like to emphasize that the mass parameter $m$
at this point is completely
arbitrary. In order to make definite predictions within SPT, 
we need a prescription for it 
as a function of $g$ and $T$.
The prescription of
Karsch, Patk\'os, and Petreczky for $m$ is the solution to the
one-loop gap equation:
\bqa\nonumber
m_{}^2={1\over2}\alpha(\mu)\left[
\int{dp\;p\over e^{\beta\sqrt{p^2+m^2}}-1}
\right.&&
\\
\left.-\left(2\log{\mu\over m}+1\right)m^2
\right]
\;,
\label{pet}
\eqa
Their choice for the renormalization scale was $\mu=T$.
In the weak-coupling limit, the solution to~(\ref{pet}) is 
$m=g(\mu)T/\sqrt{24}$. 
There are many possibilities for generalizing Eq.~(\ref{pet}) to higher orders
in the coupling constant
$g$. A thorough discussion can be found in Ref.~\cite{ABS-01}.
\hspace{3cm} 
%Below, I three different choices.
{\it Screening Mass}:\\
The screening mass is given by the pole of the propagator at zero frequency:
\bqa
p^2+\Pi({p},0)=0\;,\hspace{0.3cm}p^2=m_s^2\;.
\eqa
{\it Tadpole mass}:
\bqa
m^2&=&g^2\langle\phi^2\rangle\;.
\eqa
{\it Variational mass}:
\bqa
{d{\cal F}\over dm^2}&=&0\;.
\eqa
A few remarks are in order. In scalar theory, the screening mass can be
calculated to all orders in SPT, but in QCD it cannot be calculated beyond
leading order due to infrared divergences arising at the nonperturbative 
magnetic scale $g^2T$. The tadpole mass cannot easily be generalized to 
gauge theories, since a term $\langle A_{\mu}A_{\mu}\rangle$ is gauge variant.

The screening mass, the tadpole mass, and the variational mass 
all satisfy the same gap equation at leading order in $g$ and coincides
with the gap equation used by 
Karsch, Patk\'os, and Petreczky. At two loops, however, the gap equations
differ.
Fig.~\ref{fig2} shows the two- and three-loop approximations to the
pressure normalized to that of an ideal gas of massless bosons.
For comparison, the successive weak-coupling approximations to the pressure
are also shown (curves labelled 2 to 5). From Fig.~\ref{fig2}, it is obvious
that SPT converges much better than the weak-coupling expansion.
I have not shown the bands that one obtains by varying the renormalization
scale $\mu$ around the central value $2\pi T$ by a factor of two, but both
bands lie well within the band of the $g^5$ approximation that is shown
in Fig.~\ref{fig1}. This indicates that the uncertainty in SPT is significantly
reduced compare to the weak-coupling expansion. 
See also Ref.~\cite{ABS-01} for a thorough discussion.

\section{Hard Thermal Loop Perturbation Theory}
%Hard Thermal Loop perturbation Theory 
HTLPT is the generalization of 
SPT to gauge theories~\cite{ABS-99}. 
One cannot simply add and subtract a local mass term
as this would violate gauge invariance~\footnote{One can, however, add
a term proportional to the Polyakov loop, which would act like a mass term
for the zero-frequency component of $A_0$. I thank L.G. Yaffe for pointing
this out to me.}. 
Instead one adds and subtracts
to the QCD Lagrangian an HTL improvement term:
\bqa
{\cal L }&=&{\cal L}_{\rm QCD}+{\cal L}_{\rm HTL}+\Delta{\cal L}_{\rm HTL}\;,
\eqa
where $\Delta{\cal L}_{\rm HTL}$ includes extra counterterms 
necessary. %It is not known whether HTLPT is renormalizable
${\cal L}_{\rm HTL}$ is proportional to $(1-\delta)m_D^2$, where
$m_D$ is a variational mass parameter that can be identified with the
Debye mass.
HTLPT is then 
a systematic expansion in $\alpha_s$ and $\delta$.
The QCD pressure at leading order in HTLPT can be calculated exactly
by replacing the sum over Matsubara frequencies, extracting the
the poles in $\epsilon$, and then reducing the momentum integrals
that were at most two-dimensional and could therefore be easily 
evaluated~\cite{ABS-99,ABS-00}.
%If we tried to do this at next-to-leading order, we would have to evaluate
%integrals that were five-dimensional. 
This is intractable at NLO, and we therefore
evaluated the sum-intgrals 
approximately by expanding them in powers of $m_D/T$.
This was done at the three loops
in SPT~\cite{AS-01}, 
and it was shown that reasonable approximations are obtained
after including a few terms in this expansion.

\begin{wrapfigure}[]{l}[0pt]{8cm}
%\begin{figure}[htb]
\epsfysize=6cm
\epsffile{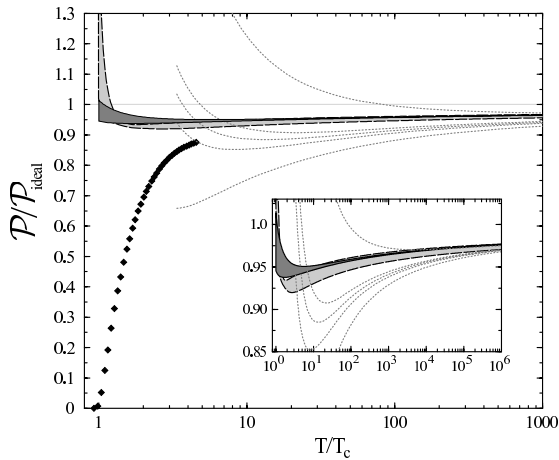}
%\vspace{2mm}
\caption[a]{
The LO and NLO results for the pressure in HTLPT
compared with 4-d lattice results (diamonds)
and 3-d lattice results (dotted lines) for various values of an unknown
coefficient in the 3-d effective Lagrangian.
The LO HTLPT result is shown as a light-shaded band outlined by a dashed line.
The NLO HTLPT result is shown as a dark-shaded band outlined by a solid line.
%The shaded bands correspond to variations
%of the renormalization scale $\mu$ by a factor of two around $\mu=2\pi T$.
}
\label{fig3}
\vspace{-1cm}
%\end{figure}
\end{wrapfigure}
The final results for the LO and NLO HTLPT predictions for the pressure
of pure-glue QCD are plotted in Fig.~\ref{fig3} as a function of $T/T_c$,
where $T_c$ is the deconfinement transition temperature~\cite{ABPS}.
The bands are again obtained by varying the renormalization scale $\mu$
by a factor of two around its central value $\mu=2\pi T$.
The two bands overlap all the way to $T_c$ and they are very narrow compared
to the corresponding bands for the weak-coupling expansion.
In Fig.~\ref{fig3}, we have also included the four-dimensional lattice
results of Boyd {\it et al}~\cite{boyd} and the three-dimensional lattice
results of Kajantie {\it et al}~\cite{KLRS}.
The LO and NLO predictions of HTLPT differ significantly from the 
four-dimensional lattice result in the whole temperature range where they
are available. In the high-temperature limit, the HTLPT pressure
approaches that of an ideal gas. This is in qualitative agreement with
the three-dimensional lattice simulations of Ref~.\cite{KLRS}.

\section{Summary}
In this talk, I have briefly discussed screened perturbation theory and
hard thermal loop perturbation theory which is a way of reorganizing the
perturbative expansion for thermal scalar field theory and 
thermal gauge theories, respectively.
Compared to the conventional weak-coupling expansion for thermodynamic
quantities such as the pressure, SPT and HTLPT
show dramatically improved convergence properties. SPT and HTLPT therefore
represent a consistent framework for calculating static and dynamical 
properties of thermal field theories.
However, the resulting predictions for e.g. the pressure fails to agree
with temperatures for which they are available.
The failure of HTLPT for temperatures below $\sim20 T_c$ could be
that a quasi-particle picture is not a good one and a description in terms
of e.g. Wilson lines~\cite{wilson} is more appropriate.

\section*{Acknowledgments}
This work was carried out in collaboration with Eric Braaten,
Emmanuel Petitgirard and Michael
Strickland.
The author would like to thank the organizers of SEWM 2002 
for a stimulating meeting. 
%J.O.A. was supported by the 
%Stichting voor Fundamenteel Onderzoek der Materie (FOM), which is 
%supported by the Nederlandse Organisatie voor Wetenschappelijk 
%Onderzoek (NWO). E.B. and E.P. were supported in part by Department of 
%Energy grant DE-FG02-91-ER4069.  M.S. was supported by US DOE Grants 
%DE-FG02-96ER40945 and DE-FG03-97ER41014.  

\end{document}